\documentclass{aastex}
\usepackage{spr-astr-addons}
\usepackage{url}

\RequirePackage{color}

\begin{document}

\title{Some non linear interactions in polytropic gas cosmology: Phase space analysis}
\shorttitle{Non linear interactions}
\shortauthors{Autors et al.}

\author{Martiros Khurshudyan\altaffilmark{1}} 
\email{khurshudyan@yandex.ru}

\altaffiltext{1}{Armenian State Pedagogical University, 375010 Yerevan, Republic of Armenia}

\begin{abstract}
There are various cosmological models with polytropic equation of state associated to dark energy. Polytropic EoS has important applications in astrophysics, therefore a study of it on cosmological framework continues to be interesting. From the other hand, there are various forms of interactions phenomenologically involved into the darkness of the universe able to solve important cosmological problems. This is a motivation for us to perform a phase space analysis of various cosmological scenarios where non linear interacting polytropic gas models are involved. Dark matter is taken to be a pressureless fluid.\end{abstract}



\section{Introduction}\label{s:INT}
Various approaches do exist to explain an accelerated expansion of the large scale universe~\citep{Riess98, Perlmutter99, Tegmark04, Abazajian04, Abazajian05, Hawkins03}. We can explain available observational data using different forms of unknown "material" known as dark energy~\citep{Jaewon12, Setare07a, Setare09a, Setare09b}~(and references therein). For instance, scalar fields can be used~\citep{Mingzhe05, Mingzhe12}. However, physical reason of the accelerated expansion of the large scale universe could be hidden in another place, therefore various other ideas also have been developed~\citep{Timothy12}, including backreaction~\citep{Villata13, Rasanen11, Buchert12}. Dark energy models can be discriminated between two groups. The first group does contain models where dark energy models have explicitly given EoS~\citep{Bento02, Bento03, Kamenshchik01, Pourhassan14, Kremer03, Capozziello05}, while the second group does contain the models, where the form of the energy density is known~\citep{RongGen11, Martiros15, Som14}. These two approaches, actually, do not reduce priority of the models, because this is mathematical trick. Recently, dark fluids have been studied intensively, due to the fact that with a given EoS, which could be non linear also, it is possible to reduce mathematical complexity of the problem. Reducing of the complexity does not mean that we should lose important solutions. In nature we have either linearity or non linearity. In case of EoS of the dark fluids, it is the same. Consideration of non linear EoS does mean, that in this way we can include some physical processes, which is not possible to do in case of a linear EoS. Viscosity, is going to be one of the actively studied options giving a rise to non linear EoS fluids~\citep{Fuhrer15, Bamba15, Acquaviva15, Giovannini15, Brevik15, Sasidharan14}. EoS of the dark energy~(dark fluid) can be a solution of an algebraic or a differential equation as well~\citep{Bamba12, Nojiri05}. Study of dynamical dark energy models has been suggested, after the discovery that the simplest model of the dark energy, cosmological constant $\Lambda$, does give the cosmological coincidence problem~\citep{Velten14}. However not all cosmological models containing a dynamical dark energy, can give results compatible to observational data. Very soon it has been found that an interaction between the components of the universe can give desirable results, namely, can solve the cosmological coincidence problem and give appropriate behavior to the cosmological parameters. Considered forms of the interaction have phenomenological origin and mainly formed from the dimensional analysis~\citep{Gua05, Wei06, Cai05, Zhang06, Wu07, Chen09, He09, Chimento10, Setare06, Setare07b, Setare07c, Setare07d, Jamil08}. It is not excluded that a fundamental theory explaining the origin of the interaction could be constructed when the nature of dark energy and dark matter will be established. Various interesting interacting dark energy models has been considered, however, still there is not any success towards the discovery of the fundamental theory. However, from the other hand observational data does support the idea of interacting dark components. Moreover, recently, using modern observational data, it was possible to show about an existence of sign changeable interactions~\citep{Cai10, Wei11, Forte14}. There is a huge number of cosmological models where EoS of the dark energy is associated to the EoS of polytropic gas~\citep{Karami09}
\begin{equation}
P_{p} = K \rho_{p}^{1+1/n},
\end{equation}
where $K$ and $n$ are two constants and $\rho_{p}$ it is the energy density of the dark energy. Interest towards such EoS is due to its applications in astrophysics~(see for instance Ref.~\citep{Karami09} and references therein). Recently, some examples of non linear interactions between dark energy and dark matter have been considered~-~an exchange between the components of the universe is a non linear process~\citep{Koyama07, Arevalo12, MKhurshudyan15}. In this paper we are interested by two phenomenological possibilities, namely, some examples of non linear and sign changeable non linear interactions are constructed. We will perform stability analysis of various cosmological models, where polytropic EoS is associated to dark energy, while dark matter is considered as a pressureless fluid. Particularly, we are interested by cosmological scenarios where into the darkness of the universe two following interactions could be involved 
\begin{equation}\label{eqn:intQ1}
Q= 3Hb \left ( \rho + \frac{\hat{\rho}}{\rho}\right ),
\end{equation}
and 
\begin{equation}\label{eqn:intQ2}
Q= 3Hbq \left ( \rho + \frac{\hat{\rho}}{\rho}\right ),
\end{equation}
where $H$ it is the Hubble parameter, $b$ is a constant, $q$ it is the deceleration parameter, $\rho$ it is the energy density either of the effective fluid or one of the components of the effective fluid. $\hat{\rho}=\rho_{i}\rho_{j}$ it is the product of energy densities of dark energy and dark matter~(if i=j we simply have $\rho_{i}^{2}$). According to very high accurate approximation, the dynamics of the large scale universe is due to dark energy and dark matter, therefore an effective dark fluid is considered, which can be described as 
\begin{equation}\label{eqn:rhoeff}
\rho_{eff} = \rho_{DM} + \rho_{DE},
\end{equation}
and 
\begin{equation}\label{eqn:Peff}
P_{eff} = P_{DM} + P_{DE}.
\end{equation}
It is known that a phase space does contain all possible states of the system. Therefore, instead to solve the field equations for some initial conditions, it is convenient to perform phase space analysis and understand the qualitative behavior of a cosmological model. In cosmology late time attractors are of great interest, because they provide relatively simple way to understand the large scale universe for a huge class of initial conditions.\\\\
The paper is organised as follows: In section~\ref{s:IntModels} we will give the definition of the interacting models in modern cosmology and will present basics of the phase space analysis to find late time attractor solutions of the field equations of General Relativity. In section~\ref{s:LTA} late time attractors are found and classified for various cosmological scenarios. To save a place we presented only late time attractor solutions among all critical points obtained from the autonomous system. Finally, discussion on obtained results are summarised in section~\ref{s:Discussion}.

\section{Interacting models and autonomous system}\label{s:IntModels}
It is well known that to describe the dynamics of the large scale flat FRW universe we need the following set of equations
\begin{equation}\label{eqn:Fridmman vlambda}
H^{2}=\frac{\dot{a}^{2}}{a^{2}}=\frac{8\pi G\rho}{3},
\end{equation}
and
\begin{equation}\label{eqn:fridman2}
\frac{\ddot{a}}{a}=-\frac{4\pi G}{3}(\rho+3P).
\end{equation}
We suppose, that the cosmological constant $\Lambda=0$, the gravitational constant $G$ and $c$ are constants with $c=8\pi G=1$. In modern cosmology an exchange between dark energy and dark matter is described in a such way that 
\begin{equation}
\dot{\rho}_{eff} + 3H (\rho_{eff} + P_{eff}) = 0,
\end{equation}
should take a place for the effective fluid. Therefore, in Physical Literature, widely accepted option describing an exchange inside the darkness is
\begin{equation}\label{eqn:firstfluid}
\dot{\rho}_{DM}+3H(\rho_{DM}+P_{DM})=Q,
\end{equation}
and
\begin{equation}\label{eqn:secondfluid}
\dot{\rho}_{DE}+3H(\rho_{DE}+P_{DE})=-Q.
\end{equation}
The form of $Q$ is determined under phenomenological assumptions, mainly, the dimensional analysis is used to construct them. It is reasonable to consider interactions which could improve previously known results and at the same time will not make the mathematical treatment of the problems complicated. It is widely believed that deeper understanding of the nature of dark energy and dark matter could give a fundamental explanation of the phenomenological assumptions about the form of $Q$. It is known that a phase space of a dynamical system it is a space in which all possible states of a system are represented. To analyse the dynamical system of the interacting polytropic gas, following to experience existing in Physical Literature~(see for instance \citep{Xu13}), we set
\begin{equation}\label{eqn:x}
x = \frac{\rho_{p}}{3H^{2}},
\end{equation}
\begin{equation}\label{eqn:y}
y = \frac{P_{p}}{3H^{2}},
\end{equation}
\begin{equation}\label{eqn:z}
z = \frac{\rho_{m}}{3H^{2}},
\end{equation}
and 
\begin{equation}\label{eqn:N}
N=\ln{a},
\end{equation}
where $a$ it is the scale factor. To have physically reasonable solutions we should have the following constraints $0 \leq x \leq 1$ and $0 \leq z \leq 1$. At the same time we should remember that $x$ and $z$ according to Eq.~(\ref{eqn:Fridmman vlambda}) should satisfy to the following constraint
\begin{equation}\label{eqn:FConst}
x+z =1.
\end{equation}
In term of $x$ and $y$ EoS parameter of the polytropic gas does read as
\begin{equation}\label{eqn:EoSC}
\omega_{p} = \frac{P_{p}}{\rho_{p}}=\frac{y}{x},
\end{equation}
while the EoS parameter of the effective fluid does read as
\begin{equation}\label{eqn:EoSeff}
\omega_{eff} = \frac{P_{p}}{\rho_{p} + \rho_{m}} = y,
\end{equation}
because dark matter is considered as a pressureless fluid. It is not hard to show that the deceleration parameter $q$ does read as
\begin{equation}\label{eqn:q}
q = -1 - \frac{\dot{H}}{H^{2}} = \frac{1}{2}(1 + 3 y).
\end{equation}
There is a huge number of articles presenting phase space analysis of different cosmological models~\citep{Xu13, Laur04, ChenXu12, Leon14, Yang11}~(to mention a few). From the next section we will start our study taking into account  general algorithm of finding critical points of the autonomous system $x^{\prime}$ and $y^{\prime}$, where $\prime$ it is the derivative with respect to $N$. Solutions of $x^{\prime}=0$ and $y^{\prime}=0$ should be found first, then the sign of the determinant and trace of the Jacobian matrix of $x^{\prime}$ and $y^{\prime}$ will point out the stability of the critical point. It is well known that if the trace of the Jacobian matrix is negative, while the determinate is positive, then the critical point is stable, because the real parts of eigenvalues are positive. From the other hand a stable critical point it is an attractor, which is we are looking for. Therefore, we need to find the range of the model parameters such, that to have a physically reasonable stable critical points i.e.  $0 \leq x_{c} \leq 1$ and $0 \leq z_{c} \leq 1$. 

\section{Late time attractors}\label{s:LTA}
In this section we will present late time attractors~(to save a place) of the field equations corresponding to several interacting polytropic dark energy models. In order to be able to perform the study we assumed the form of interaction term $Q$ to be given.

\subsection{Interaction $1$}\label{ss:Q1}
The first model which we would like to analyse is described by the following interaction
\begin{equation}\label{eqn:FormQ1}
Q = 3 H b \left ( \rho_{p} + \rho_{m} + \frac{\rho_{m}^{2}} { \rho_{p} + \rho_{m}} \right ).
\end{equation}
The autonomous system of this model does read as
\begin{equation}\label{eqn:xpQ1}
x^{\prime} = -3 b \left ( 2 + (x-2) x \right ) - 3 (1-x) y,
\end{equation}
and 
\begin{equation}\label{eqn:ypQ1}
y^{\prime} = 3y (1+y) - \frac{3(1+n)y \left ( x + y + b (2+ (x-2) x) \right )}{nx}.
\end{equation}
In this case only one physically reasonable stable critical point~$(E.1.1)$ does exist 
\begin{equation}
x = \frac{2b -1 + \sqrt{1-4b^{2}}}{2b}~~~~and~~~~y=-1,
\end{equation}
when 
\begin{equation}
n \leq -1~~~~and~~~~0< b < \frac{1}{4}\sqrt {\frac{2n^{2} - 1 }{n^{2}}} - \frac{1}{4n},
\end{equation}
or
\begin{equation}
-1 < n < 0~~~~and~~~~0< b < \frac{1}{2}.
\end{equation}
This solution is a late time scaling attractor, because 
\begin{equation}
r = \frac{\Omega_{m}}{\Omega_{p}} = \frac{\sqrt{1-4b^{2}}}{2(1-2b^{2})} - \frac{1}{2},
\end{equation}
is a constant. Our future analysis does show, that this late time scaling attractor does describe the state of the universe where EoS of the effective fluid is $\omega_{eff} = -1$, the deceleration parameter $q=-1$ and EoS of the polytropic gas does exhibit a phantom behavior (quintessence behavior is not possible) with
\begin{equation}
\omega_{p} = -\frac{2b}{2b + \sqrt{1-4b^{2}}-1},
\end{equation}
when 
\begin{equation}
n \leq -1~~~~and~~~~0< b \leq \frac{1}{2\sqrt{2}},
\end{equation}
or
\begin{equation}
-1 < n < 0~~~~and~~~~0< b < \frac{1}{2},
\end{equation}
or
\begin{equation}
-\sqrt{ \frac{1- 4 b^{2}}{ (1-8b^{2})^{2}} } + \frac{2b}{1-8b^{2}} < n \leq -1~and~\frac{1}{2\sqrt{2}}< b < \frac{1}{2},
\end{equation}

\subsection{Interaction 2}\label{ss:Q2}
The second model interesting for us is a model, where polytropic dark energy is interacting with dark matter via
\begin{equation}\label{eqn:FormQ2}
Q = 3 H b \left ( \rho_{p} + \rho_{m} + \frac{\rho_{p}^{2}} { \rho_{p} + \rho_{m}} \right ).
\end{equation}
Only one late time scaling attractor does exist for this model and it is does read as~($E.1.2$)
\begin{equation}
x = \frac{\sqrt{1-4(b-1)b} - 1}{2b},
\end{equation}
and 
\begin{equation}
y=-1.
\end{equation}
This solution is obtained from the autonomous system
\begin{equation}\label{eqn:xpQ1}
x^{\prime} = -3 b (1+x^{2})- 3(1-x)y,
\end{equation}
and 
\begin{equation}\label{eqn:ypQ1}
y^{\prime} = 3y (1+y) - \frac{3(1+n)y \left ( b + x(1+bx) + y \right )}{nx}.
\end{equation}
It is not hard to see that $\omega_{eff} =-1$, $q =-1$, while
\begin{equation}
\omega_{p} = - \frac{2b}{\sqrt{1-4(b-1)b}-1},
\end{equation}
and
\begin{equation}
r = \frac{\Omega_{m}}{\Omega_{p}} = \frac{2b + \sqrt{1-4(b-1)b}}{2(1-b)}. 
\end{equation}
Discussed late time scaling attractor does represent the universe where polytropic gas is a phantom energy, when one of the following conditions does take a place
\begin{equation}
 n \leq -1~and~0<b<\frac{n-1}{2n},
\end{equation}
or
\begin{equation}
-1 < n < 0~~~~and~~~~0<b<1.
\end{equation}

\subsection{Interaction 3}\label{ss:Q3}
In this subsection we will consider a cosmological model where interaction of the following form is involved into the darkness of the large scale universe
\begin{equation}\label{eqn:FormQ3}
Q = 3 H b \left ( \rho_{p} + \rho_{m} + \frac{\rho_{p}\rho_{m}} { \rho_{p} + \rho_{m}} \right ).
\end{equation}
Among four critical points obtained as solutions from the appropriate autonomous system only one is a physical reasonable. It is of the following form~(E.1.3)
\begin{equation}
x=\frac{1+b - \sqrt{1+(5b -2)b}}{2b},
\end{equation}
and
\begin{equation}
y = -1.
\end{equation}
Our analysis does show that this solution is late time scaling attractor describing a universe where polytropic gas is a phantom dark energy, when
\begin{equation}
n \leq -1~and~0<b<\frac{2n-5}{10n} + \frac{1}{10}\sqrt{\frac{5+4n^{2}}{n^{2}}},
\end{equation}
or
\begin{equation}
-1<n<0~~~~and~~~~0<b<1.
\end{equation}
For this model $\omega_{eff} =-1$, $q=-1$,
\begin{equation}
\omega_{p} = - \frac{2b}{1+b - \sqrt{1+(5b-2)b}},
\end{equation}
and
\begin{equation}
r =\frac{\Omega_{m}}{\Omega_{p}} = \frac{3b + \sqrt{1+(5b -2)b} -1}{2(1-b)}.
\end{equation}

\begin{figure}[tb]
 \begin{center}$
 \begin{array}{cccc}
 \includegraphics[width=80 mm]{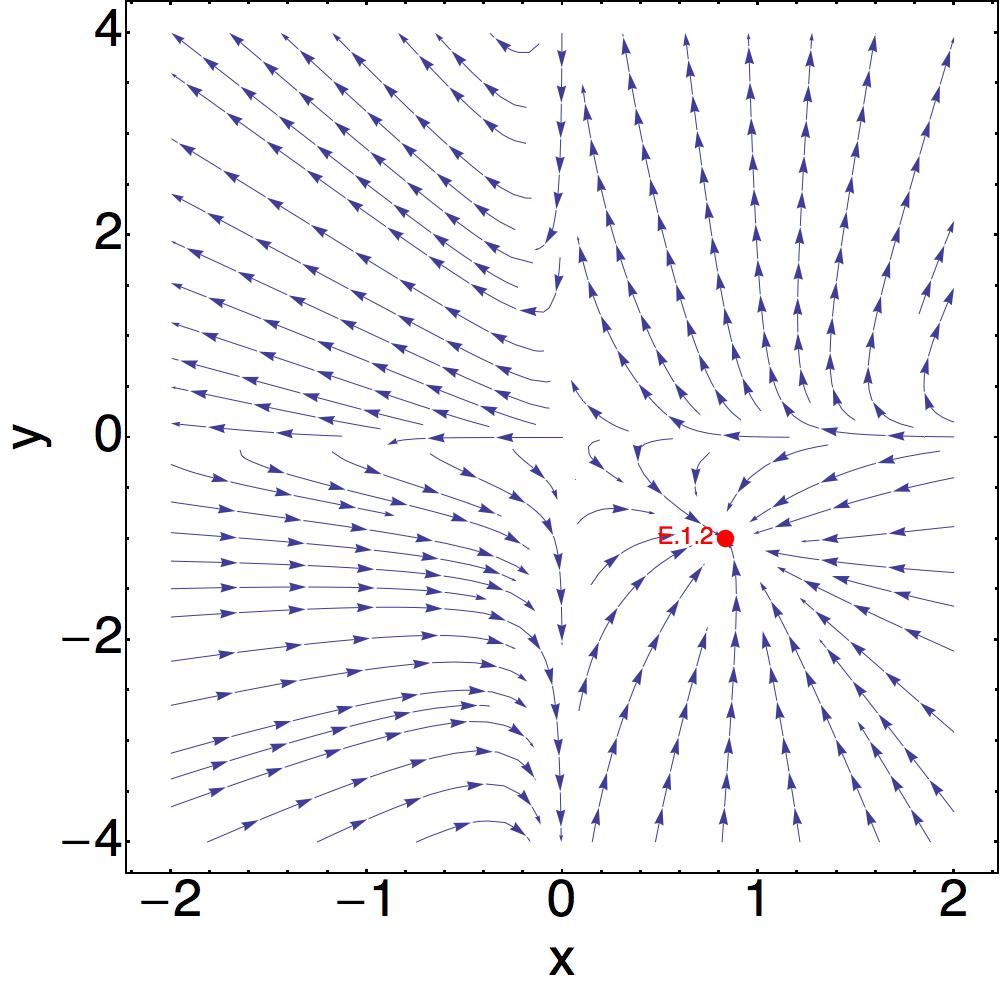} \\
\includegraphics[width=80 mm]{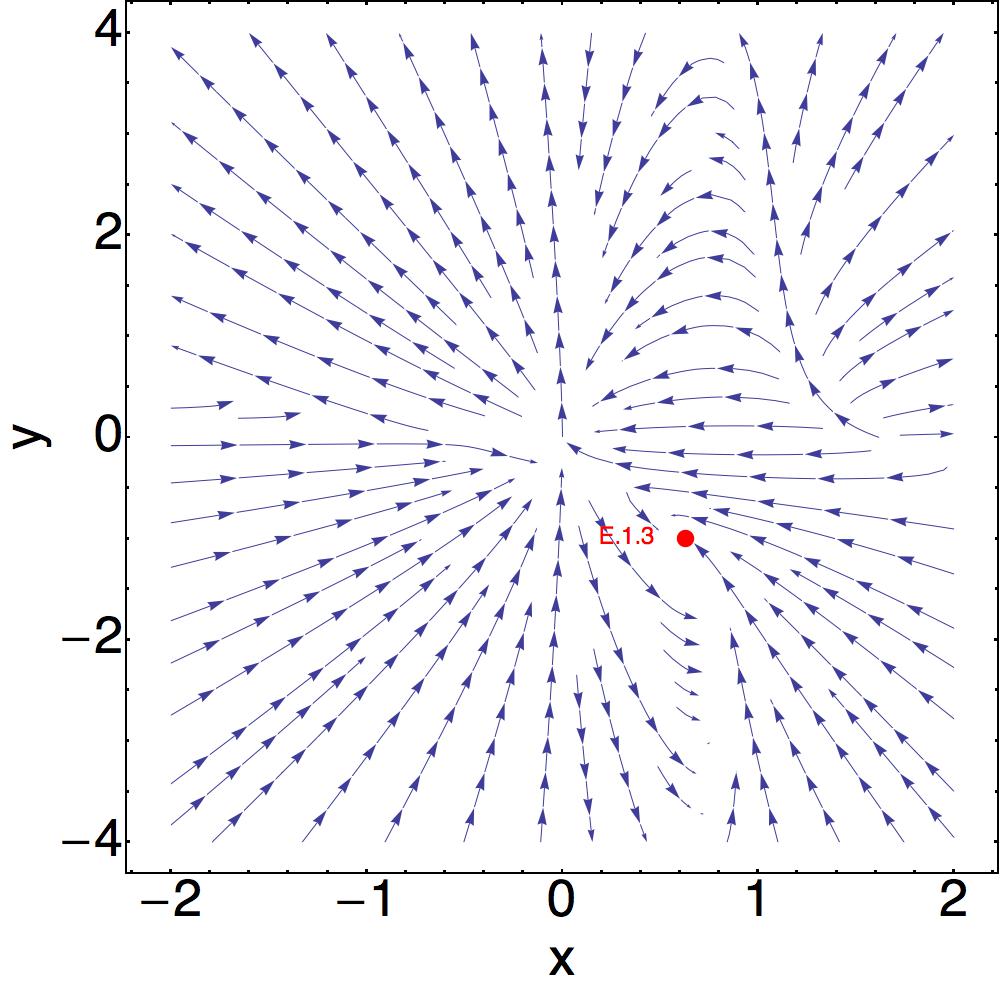} \\
 \end{array}$
 \end{center}
 \caption{Phase space portraits for the models with interaction term given via Eq.~(\ref{eqn:FormQ2}) and Eq.~(\ref{eqn:FormQ3}). For symmetry we plotted phase space portrait when $x \in [-1,1]$, however, physically reasonable range for $x\in [0,1]$}
 \label{fig:Fig1}
\end{figure}

\subsection{Sign changeable interactions}\label{ss:SCInter}
Interaction terms considered in subsections~\ref{ss:Q1}~-~\ref{ss:Q3} are interactions with fixed sign and indicate an exchange from dark energy to dark matter. Recently it was found that during the evolution of the universe interaction has changed the sign. This mean that an exchange from dark matter to dark energy has been in the history of the universe. Sign changeable interactions considered in this subsection are constructed according to well known algorithm, the deceleration parameter $q$ is involved, from the interactions considered in subsections~\ref{ss:Q1}~-~\ref{ss:Q3}. Appropriate real critical points are summarised into Table~{\ref{tbl:Table1}}. We have considered the following forms for $Q$ 
\begin{equation}\label{eqn:FormqQ1}
Q = 3 H b q \left ( \rho_{p} + \rho_{m} + \frac{\rho_{m}^{2}}{\rho_{p} + \rho_{m} } \right ),
\end{equation}
\begin{equation}\label{eqn:FormqQ2}
Q = 3 H b q \left ( \rho_{p} + \rho_{m} + \frac{\rho_{p}^{2}}{\rho_{p} + \rho_{m} } \right ),
\end{equation}
and
\begin{equation}\label{eqn:FormqQ3}
Q = 3 H b q \left ( \rho_{p} + \rho_{m} + \frac{\rho_{p}\rho_{m}}{\rho_{p} + \rho_{m} } \right ).
\end{equation}
Our analysis did show that obtained critical points are not stable. Compared to the interactions terms in subsections~\ref{ss:Q1}~-~\ref{ss:Q3}, appropriate sign changeable interactions do not allow us to obtain late time attractors.

\begin{table*}[tb]
 \small
 \caption{Critical points for the models where sign changeable non linear interactions are given via Eq.-s~(\ref{eqn:FormqQ1})~-~(\ref{eqn:FormqQ3})} 
  \label{tbl:Table1}
 \begin{tabular}{| l | l | l | l | l | l | l |}
 \tableline  
  $S. P.$ &$Q$ & $x$ & $y$ \\
 \tableline 
  $E.2.1$ & $Eq.~(\ref{eqn:FormqQ1})$  & $\frac{1+2b \pm \sqrt{1-4b^{2}}}{2b}$ & $-1$ \\
   \tableline 
   $E.2.2$  & $ Eq.~(\ref{eqn:FormqQ2}) $ & $\frac{1\pm \sqrt{1- 4b(1+b)}}{2b}$ & $-1$ \\ 
     \tableline 
   $E.2.3$  &  $ Eq.~(\ref{eqn:FormqQ3}) $ & $\frac{1}{2} (1\pm \sqrt{5})$ & $0$ \\ 
      \tableline 
   $E.2.4$  &  $ Eq.~(\ref{eqn:FormqQ3})$ & $\frac{b-1 \pm \sqrt{1+(2+5b)b}}{2b}$ & $-1$ \\
          \tableline 
 \end{tabular}
 \end{table*}

\section{Discussion}\label{s:Discussion}
In this paper various interacting polytropic gas cosmological scenarios have been considered. Instead to solve the system of differential equations for some initial conditions, we performed phase space analysis. Phase space does contain all possible states of the system, therefore does provide qualitative understanding of the cosmological scenarios. Some examples of non linear interactions involved into the darkness of the universe have been considered. Appropriate critical points and late time attractors have been found. Considered three forms of non linear interactions provide late time scaling attractors corresponding to the state, when polytropic gas is a phantom energy. Motivated by the fact that using the deceleration parameter $q$ we can construct sign changeable interactions, we considered three cosmological models with sign changeable non linear interactions. New interaction terms are constructed from non linear interactions considered in subsections~\ref{ss:Q1}~-~\ref{ss:Q3} using the deceleration parameter $q$. It was easy to obtain the critical points of appropriate autonomous systems and see that they are not stable. Which does mean that late time attractors for these models are missing. It is of a great interest to consider other forms of non linear interactions obtained from Eq.~(\ref{eqn:intQ1}) and Eq.~(\ref{eqn:intQ2}) for more comprehensive understanding of late time behavior of appropriate cosmological models. In near future, our goal is to pay a particular attention to different forms of sign changeable interactions considered in this paper.

\acknowledgments
M. Khurshudyan is grateful to Prof E. Kokanyan for warm hospitality, comprehensive support and stimulating discussion.


\begin{thebibliography}{}

\bibitem[\protect\citeauthoryear{Abazajian,~K. et al.}{2005}]{Abazajian05}
Abazajian,~K. et al.: \aj \textbf{129}, 1755-1759 (2005)

\bibitem[\protect\citeauthoryear{Abazajian,~K. et al.}{2004}]{Abazajian04}
Abazajian,~K. et al.: \aj \textbf{128}, 502-512 (2004) 

\bibitem[\protect\citeauthoryear{Acquaviva,~G. et al.}{2015}]{Acquaviva15}
Acquaviva,~G. et al.: arXiv:1505.01965 (2015)

\bibitem[\protect\citeauthoryear{Arevalo,~F. et al.}{2012}]{Arevalo12}
Arevalo,~F. et al.:  Class. Quant. Grav. \textbf{29}, 23 (2012)

\bibitem[\protect\citeauthoryear{Bamba,~K. et al.}{2012}]{Bamba12}
Bamba,~K. et al.: \apss \textbf{342}, 155-228 (2012)

\bibitem[\protect\citeauthoryear{Bamba,~K. et al.}{2015}]{Bamba15}
Bamba,~K. et al.: arXiv:1508.05451 (2015)

\bibitem[\protect\citeauthoryear{Bento,~M. C. et al.}{2002}]{Bento02}
Bento,~M. C. et al.: \prd \textbf{66}, 043507 (2002)

\bibitem[\protect\citeauthoryear{Bento,~M. C. et al.}{2012}]{Bento03}
Bento,~M. C. et al.: Gen. Rel. Grav. \textbf{35}, 2063-2069 (2003)

\bibitem[\protect\citeauthoryear{Brevik,~I. et al.}{2015}]{Brevik15}
Brevik,~I. et al.: Universe \textbf{1}, 24-37 (2015)

\bibitem[\protect\citeauthoryear{Buchert,~T. et al.}{2012}]{Buchert12}
Buchert,~T. et al.: Annu.Rev. of Nuc. and Particle Sci. \textbf{62}, 57-79 (2012)

\bibitem[\protect\citeauthoryear{Cai,~R. G. et al.}{2005}]{Cai05}
Cai,~R. G. et al.: \jcap \textbf{0503}, 002 (2005)

\bibitem[\protect\citeauthoryear{Cai,~R. G. et al.}{2010}]{Cai10}
Cai,~R. G. et al.: \prd \textbf{81}, 103514 (2010)

\bibitem[\protect\citeauthoryear{Cai,~R.-G. et al.}{2011}]{RongGen11}
Cai,~R.-G. et al.: \prd \textbf{84}, 123501 (2011)

\bibitem[\protect\citeauthoryear{Capozziello,~S. et al.}{2005}]{Capozziello05}
Capozziello,~S. et al.: \jcap \textbf{04}, 005 (2005)

\bibitem[\protect\citeauthoryear{Chen,~X. M. et al.}{2009}]{Chen09}
Chen,~X. M. et al.: \jcap \textbf{0904}, 001 (2009)

\bibitem[\protect\citeauthoryear{Chen,~X. et al.}{2012}]{ChenXu12}
Chen,~X. et al.: \jcap  \textbf{07}, 005 (2012)

\bibitem[\protect\citeauthoryear{Chimento,~L.P.}{2009}]{Chimento10}
Chimento,~L.P.: \prd \textbf{81}, 043525 (2010)

\bibitem[\protect\citeauthoryear{Clifton,~T. et al.}{2012}]{Timothy12}
Clifton,~T. et al.: Physics Reports \textbf{513}, 1 (2012)

\bibitem[\protect\citeauthoryear{Forte,~M.}{2014}]{Forte14}
Forte,~M.: Gen. Rel. Grav. \textbf{46}, 1811 (2014)

\bibitem[\protect\citeauthoryear{Fuhrer,~F. et al.}{2015}]{Fuhrer15}
Fuhrer,~F. et al.: arXiv:1509.03073 (2015)

\bibitem[\protect\citeauthoryear{Giovannini,~M.}{2015}]{Giovannini15}
Giovannini,~M.: Class. Quant. Grav. \textbf{32}, 15 (2015)

\bibitem[\protect\citeauthoryear{Guo,~Z. K. et al.}{2005}]{Gua05}
Guo,~K. et al.: \jcap \textbf{0505}, 002 (2005)

\bibitem[\protect\citeauthoryear{Hawkins,~E. et al.}{2003}]{Hawkins03}
Hawkins,~E. et al.: \mnras \textbf{346}, 78-96 (2003)

\bibitem[\protect\citeauthoryear{He,~J. H. et al.}{2009}]{He09}
He,~J. H. et al.: \prd \textbf{80}, 063530 (2009)

\bibitem[\protect\citeauthoryear{Jamil,~M. et al.}{2008}]{Jamil08}
Jamil,~M. et al.: Eur. Phys. J. C \textbf{56}, 429 (2008)

\bibitem[\protect\citeauthoryear{Jarv,~L. et al.}{2004}]{Laur04}
Jarv,~L. et al.: \jcap \textbf{0408}, 016 (2004)  

\bibitem[\protect\citeauthoryear{Kamenshchik,~A. Yu. et al.}{2012}]{Kamenshchik01}
Kamenshchik,~A. Yu. et al.: Phys. Lett. B \textbf{511}. 265-268 (2001) 

\bibitem[\protect\citeauthoryear{Karami,~K. et al.}{2009}]{Karami09}
Karami,~K. et al.: Eur. Phys. J. C \textbf{64}, 85 (2009)

\bibitem[\protect\citeauthoryear{Khurshudyan,~M. et al.}{2015}]{Martiros15}
Khurshudyan,~M. et al.: \apss \textbf{357}, 113 (2015)

\bibitem[\protect\citeauthoryear{Khurshudyan,~M. et al.}{2015}]{MKhurshudyan15}
Khurshudyan,~M. et al.: arXiv:1509.02263 (2015)

\bibitem[\protect\citeauthoryear{Kremer,~G. M.}{2003}]{Kremer03}
Kremer,~G. M.: \prd \textbf{68}, 123507 (2003)

\bibitem[\protect\citeauthoryear{Koyama,~K. et al.}{2007}]{Koyama07}
Koyama,~K. et al.: \prd \textbf{75}, 084040 (2007)

\bibitem[\protect\citeauthoryear{Leon,~G. et al.}{2014}]{Leon14}
Leon,~G. et al.: Phys. Lett. B \textbf{732}, 285 (2014)

\bibitem[\protect\citeauthoryear{Li,~M. et al.}{2012}]{Mingzhe12}
Li,~M. et al.: \jcap \textbf{04}, 003 (2012)

\bibitem[\protect\citeauthoryear{Li,~M. et al.}{2005}]{Mingzhe05}
Li,~M.: \jcap \textbf{12}, 002 (2005)

\bibitem[\protect\citeauthoryear{Nojiri,~S. et al.}{2005}]{Nojiri05}
Nojiri,~S. et al.: \prd{72}, 023003 (2005)

\bibitem[\protect\citeauthoryear{Perlmutter,~S. et al.}{1999}]{Perlmutter99}
Perlmutter,~S. et al.: \apj \textbf{517}, 565-586 (1999)

\bibitem[\protect\citeauthoryear{Pourhassan,~B. et al.}{2014}]{Pourhassan14}
Pourhassan,~B. et al.: Results in Physics 4, 101-102 (2014) 

\bibitem[\protect\citeauthoryear{Rasanen,~S. et al.}{2011}]{Rasanen11}
Rasanen,~S. et al.: Class. Quantum Grav. \textbf{28}, 164008 (2011)

\bibitem[\protect\citeauthoryear{Riess,~A.G et al.}{1998}]{Riess98}
Riess,~A.G et al.: \aj\ \textbf{116}, 1009 (1998)

\bibitem[\protect\citeauthoryear{Sasidharan,~A. et al.}{2014}]{Sasidharan14}
Sasidharan,~A. et al.: arXiv:1411.5154 (2014)

\bibitem[\protect\citeauthoryear{Setare,~M. R.}{2006}]{Setare06}
Setare,~M. R.: Phys. Lett. B \textbf{642}, 421 (2006)

\bibitem[\protect\citeauthoryear{Setare,~M. R.}{2007a}]{Setare07a}
Setare,~M.R.: Phys. Lett. B \textbf{653}, 116 (2007)

\bibitem[\protect\citeauthoryear{Setare,~M. R.}{2007b}]{Setare07b}
Setare,~M.R.: Phys. Lett. B \textbf{648}, 329 (2007)

\bibitem[\protect\citeauthoryear{Setare,~M. R.}{2007c}]{Setare07c}
Setare,~M.R.: Eur. Phys. J. C \textbf{52}, 689 (2007)

\bibitem[\protect\citeauthoryear{Setare,~M. R.}{2007d}]{Setare07d}
Setare,~M.R.: Phys, Lett B \textbf{654}, 1 (2007)

\bibitem[\protect\citeauthoryear{Setare,~M. R.}{2009a}]{Setare09a}
Setare~M. R. et al.: Phys.Lett.B \textbf{673}, 241 (2009)

\bibitem[\protect\citeauthoryear{Setare,~M. R.}{2009b}]{Setare09b}
Setare~M. R.:  Int. J. Mod. Phys. D \textbf{18}, 419 (2009)

\bibitem[\protect\citeauthoryear{Som,~S. et al.}{2014}]{Som14}
Som,~S. et al.: \apss \textbf{352}, 867 (2014)

\bibitem[\protect\citeauthoryear{Tegmark,~M. et al.}{2004}]{Tegmark04}
Tegmark,~M. et al.: \prd \textbf{69}, 103501 (2004)

\bibitem[\protect\citeauthoryear{Velten,~H et al.}{2014}]{Velten14} 
Velten,~H et al.: Eur. Phys. J. C \textbf{74}, 11 (2014) 

\bibitem[\protect\citeauthoryear{Villata,~M.}{2013}]{Villata13}
Villata,~M.: \apss \textbf{345}, 1-9 (2013)

\bibitem[\protect\citeauthoryear{Wei,~H. et al.}{2006}]{Wei06}
Wei,~H. et al.: \prd \textbf{73}, 083002 (2006)

\bibitem[\protect\citeauthoryear{Wei,~H.}{2011}]{Wei11}
Wei,~H.: Nucl. Phys. B \textbf{845}, 381 (2011)

\bibitem[\protect\citeauthoryear{Wu,~P. et al.}{2007}]{Wu07}
Wu,~P. et al.: Class. Quantum Gravity \textbf{24}, 4661 (2007)

\bibitem[\protect\citeauthoryear{Xu,~Y. D. et al.}{2013}]{Xu13}
Xu,~Y. D. et al.: \apss \textbf {343}, 807  (2013)

\bibitem[\protect\citeauthoryear{Yang,~R. et al.}{2011}]{Yang11}
Yang,~R. et al.: Class. Quant. Grav. \textbf{28}, 065012 (2011)

\bibitem[\protect\citeauthoryear{Yoo,~J. et al.}{2003}]{Jaewon12}
Yoo,~J. et al.: Int. J. Mod. Phys. D \textbf{21}, 1230002 (2012)

\bibitem[\protect\citeauthoryear{Zhang,~H. et al.}{2005}]{Zhang06}
Zhang,~H. et al.: \prd {73}, 043518 (2006)

\end{thebibliography}

\end{document}